\documentclass[conference]{IEEEtran}

\usepackage{graphics, array, url, hyperref, epsfig, amsmath, amssymb, graphicx}
\usepackage{subfigure}
\usepackage{threeparttable}
\usepackage{extarrows}
\usepackage{color}

\begin{document}

\title{Content Protection in Named Data Networking: Challenges and Potential Solutions}

\author{\IEEEauthorblockN{Yong Yu$^{\dagger}$, Yannan Li$^{\ddagger}$, Xiaojiang Du$^{\S}$, Ruonan Chen$^{\dagger}$ and Bo Yang$^{\dagger}$}
 \IEEEauthorblockA{$^{\dagger} $School of Computer Science, Shaanxi Normal University, Xi'an 710062, China.
\\ Email: \{yuyong, chenruonan, byang\}@snnu.edu.cn}
\IEEEauthorblockA{$^{\ddagger}$ School of Computing and Information Technology, University of Wollongong, Wollongong,\\ NSW 2522, Australia. Email: liyannan2016@163.com (Corresponding Author)}
\IEEEauthorblockA{$^{\S}$Dept. of Computer and Information Sciences, Temple University, Philadelphia, PA, 19122, USA.  \\Email: dxj@ieee.org (Corresponding Author)}
}

\maketitle

\begin{abstract}
Information-Centric Networks (ICN) are promising alternatives to current Internet architecture since the Internet struggles with a number of issues such as scalability, mobility and security. ICN offers a number of potential benefits including reduced congestion and enhanced delivery performance by employing content caching, simpler network configurations and stronger security for the content. Named Data Networking (NDN), an instance of the ICN, enables content delivery instead of host-centric approaches by naming data rather than host. In order to make NDN practical in the real world, the challenging issues of content security need to be addressed.
In this article, we examine the architecture, content security as well as possible solutions to these issues of NDN, with a special focus on the content integrity and provenance. We propose a variety of digital signature schemes to achieve the data integrity and origin authentication in NDN for various applications, which include cost-effective signatures,
privacy preserving signatures,
network coding signatures, and post-quantum signatures. We also present the speed-up techniques in generating signatures and verifying signatures such as pre-computation, batch verification and server-aided verification to reduce the computational cost of the producers and receivers in NDN. A number of certificate-free trust management approaches and possible adoptions in NDN are investigated.


%
%

\vskip 2mm \noindent{\bf Keywords:}\quad Information-Centric Networks, Named Data Networking, Content Integrity, Digital Signatures.
\end{abstract}

\section{Introduction}

Internet, a packet data network in which clients and servers with a specific IP address can interact over a pre-established communication channel for conversion, has achieved great success since its invention 36 years ago. Internet users can transfer various content such as audio, text, image and video packets over Internet thanks to the well-known TCP/IP protocol. The current Internet's hourglass architecture focuses on IP network layer, in which the minimal functionality required for global interconnectivity is implemented. However, profound changes in nature of Internet communication have taken place where massive content are produced and distributed over the Internet. An increasing number of applications involving data delivery are concerned more about what data are required than where the data are from. Moreover, there is no in-built mechanisms supporting mobility and security in Internet. Thus, it is fair to say that the Internet architecture poorly matches its primary use today in the sense that scalable content distribution is poorly served by the Internet.

The drawbacks of Internet mentioned above motivated people to find a promising alternative supporting content-centric communication to the Internet. Five projects for content-based Internet paradigms \cite{icn} under its Future Internet Architecture Program were funded by NSF, and Named Data Networking (NDN) \cite{ndn} is one of the projects.
  Different from IP, which centers on communication end-points via their IP addresses, NDN does not utilize source and destination addresses but makes use of data packets with content names. NDN places data as the highest priority by naming them instead of their addresses.  The employment of content names in NDN communication enables routers to monitor the states of the packets, which provides additional functions the IP routers cannot offer. The NDN packets are independent of their location, which supports in-network caching of contents for future use and supports consumer with different positions inherently.

\begin{figure}[h]
  \centering
  \includegraphics[width=0.4\textwidth]{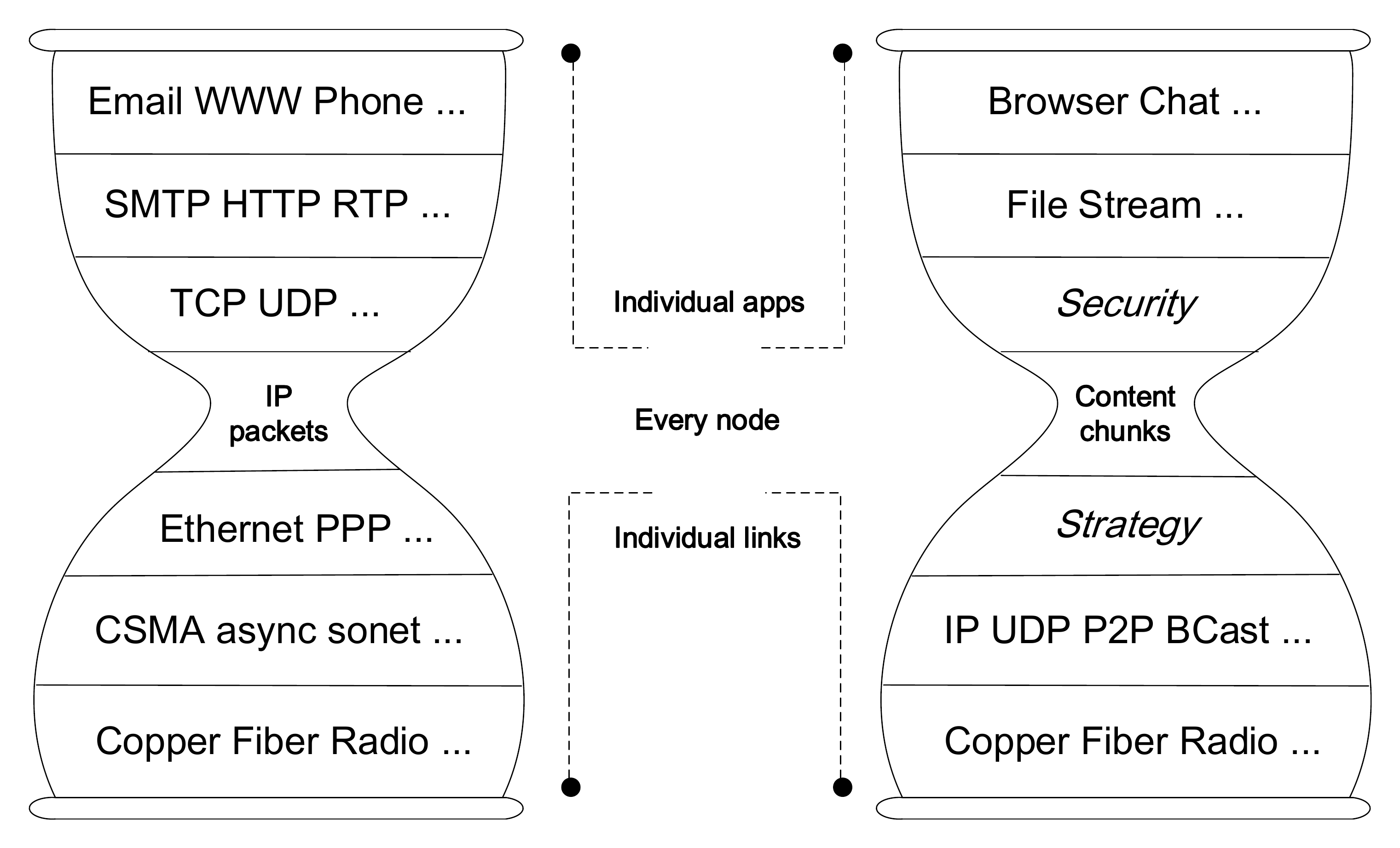}\\
  \caption{Internet and NDN Hourglass Architectures}\label{fig:hourglass}
\end{figure}

In the NDN project(named-data.net), NDN was described with the similar layered hourglass architecture \cite{ndn} as that of the Internet except a few different functions between corresponding network layers as shown in Fig. \ref{fig:hourglass}. To be more specific, firstly, the network service in the Internet is delivering packets to a specified destination IP address while it reteives data with a unique defined name in NDN. The main overlays in NDN are named content chunks while the fundamental element of Internet is a channel between two endpoints with a unique IP address.
Secondly, there are two new layers named security and strategy in NDN protocol stack. Security layer guarantees the security of content, rather than the whole communication channel, while strategy layer is for NDN forwarding plane, where the functions of Internet's transport layer are embeded, thus no need for some separate transport layers in NDN.

\section{Architecture of NDN}

NDN adopts a hierarchial structure \cite{ndn}, which is user friendly, to construct content names. Each name might consist of multiple parts: source of the content, the file name, version of the content, $e.g.$, we can name segment 1 of version 1 of a SNNU picture like $ /snnu/images/a.jpg/v1/s1$, where the sign "/" denotes the boundary of each part of the name. With the help of hierarchial namespaces and naming conventions, a consumer and a provider can always construct the same content name in this hierarchial structure, thus NDN can fetch desirable content by names.

\begin{figure}[h]
  \centering
  \includegraphics[width=0.4\textwidth]{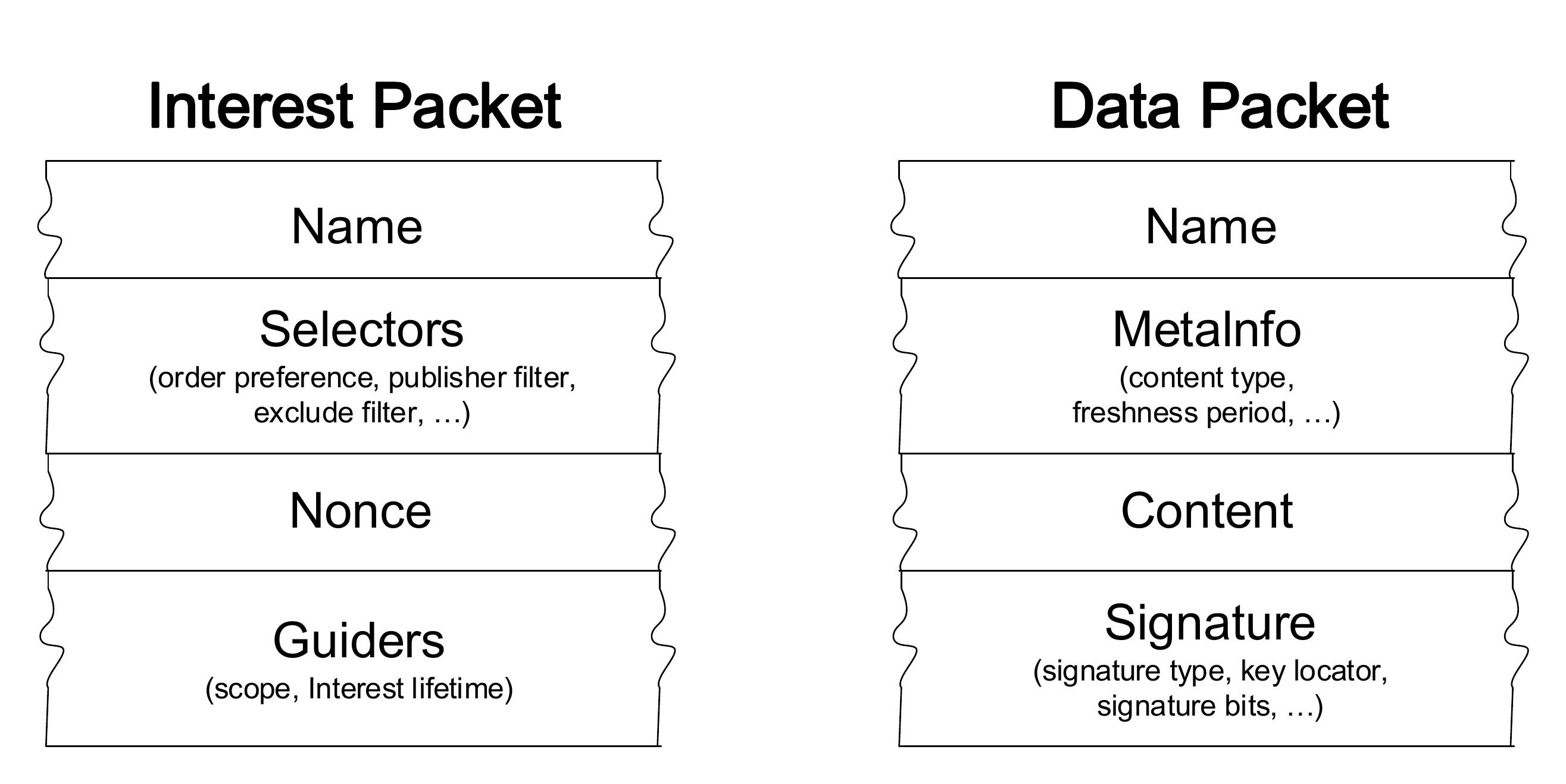}\\
  \caption{Packets in the NDN Architecture}\label{fig:packet}
\end{figure}

\begin{figure}[h]
  \centering
  \includegraphics[width=0.5\textwidth]{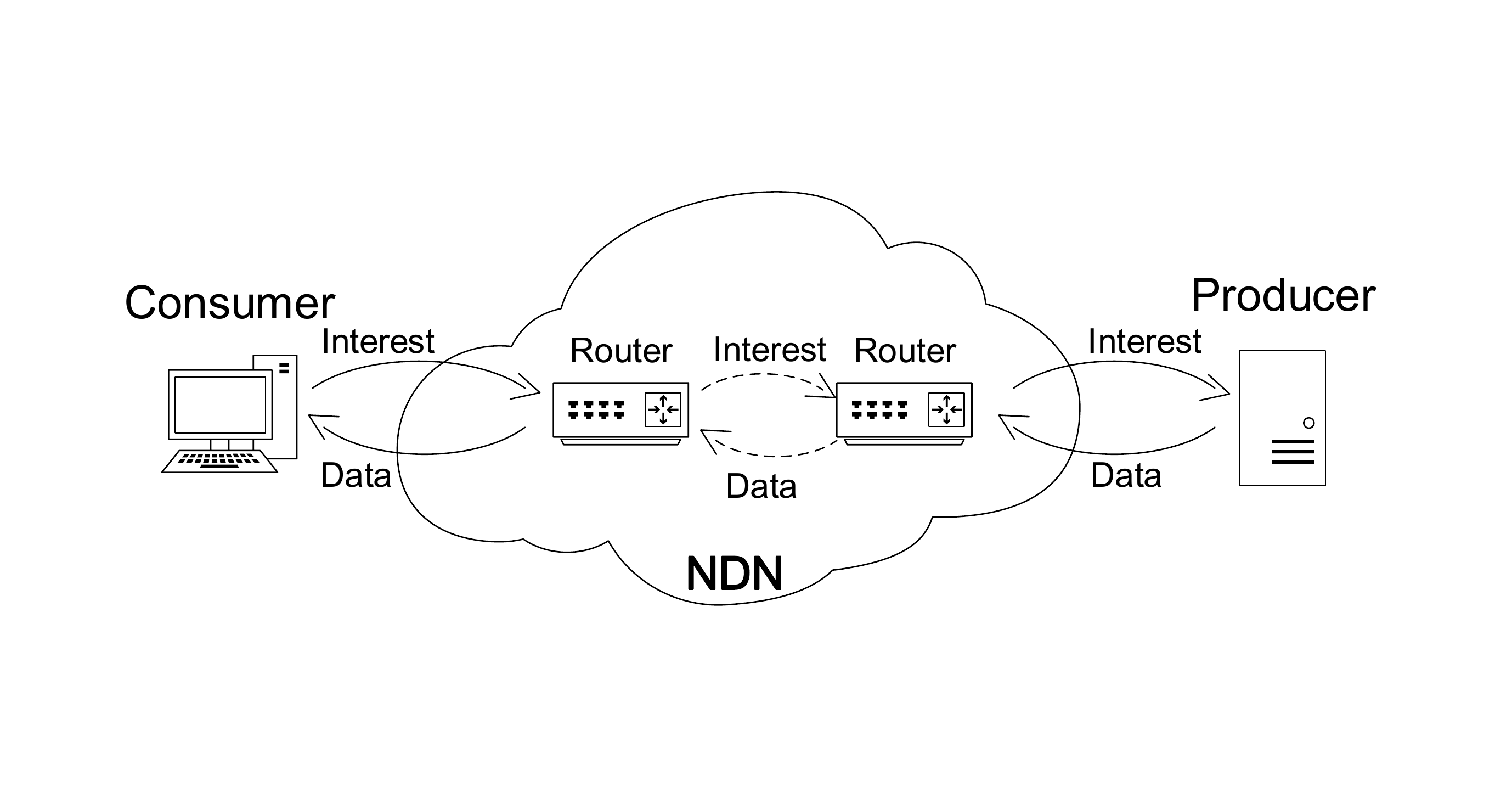}\\
  \caption{An Abbreviated Version of NDN}\label{fig:process}
\end{figure}

 Communication in NDN \cite{ndn} involves two types of packets: $Interest$ and $Data$ as well as three types of roles: $Consumer$, $Provider$ and $Router$. As indicated in Fig. \ref{fig:packet}, content names are embedded in both packets to ensure that each Interest package has one and only one Data package paired with it. An abbreviated process shown in Fig. \ref{fig:process} tells a round of content search and retrieval. The $Consumer$ issues an $Interest$ packet including the name of a desired content and sends the packet to the network. Then the $Interest$ packet will travel through NDN $Routers$ which own the capacity of caching and forwarding packets. When the $Interest$ packet arrives at the node which indeed generates the requested content, this node, termed as $Producer$, returns a $Data$ packet containing the consistent name and the specific content. The $Data$ packet just reverses the incoming path of the $Interest$ packet to reach the $Consumer$.

The data structures maintained by NDN routers include a $Content$ $Store$ (CS), a $Pending$ $Interest$ $Table$ (PIT) and a $Forwarding$ $Information$ $Base$ (FIB). CS stores data packets passing through a router. Because of the special packet structure, the data can be cached for future reuse. Once the \emph{Interest} reaches a router that caches the required data, it will return a data packet to the consumer immediately. Owing to the limitation of cache size, these data packets will be dropped directly (lifetime expires) or replaced by new data packets (cache is full) according to the caching policy. PIT maintains a record for the data name carried in \emph{Interest} packet as well as its incoming faces. If the router receives a \emph{Data} packet, but there is no record about its name in PIT, the router will drop the packet directly. FIB maintains an entry for name prefix and outgoing faces. When the content name doesn't exist in CS and PIT, FIB will forward the \emph{Interest} to the corresponding next-hops. In extreme cases, there is no prefix-name matches the \emph{Interest} packet, FIB may trigger the router to explore potential alternate paths. $Forwarding$ $Strategy$ model makes these three parts work closely together.

\begin{figure}[h]
  \centering
  \includegraphics[width=0.4\textwidth]{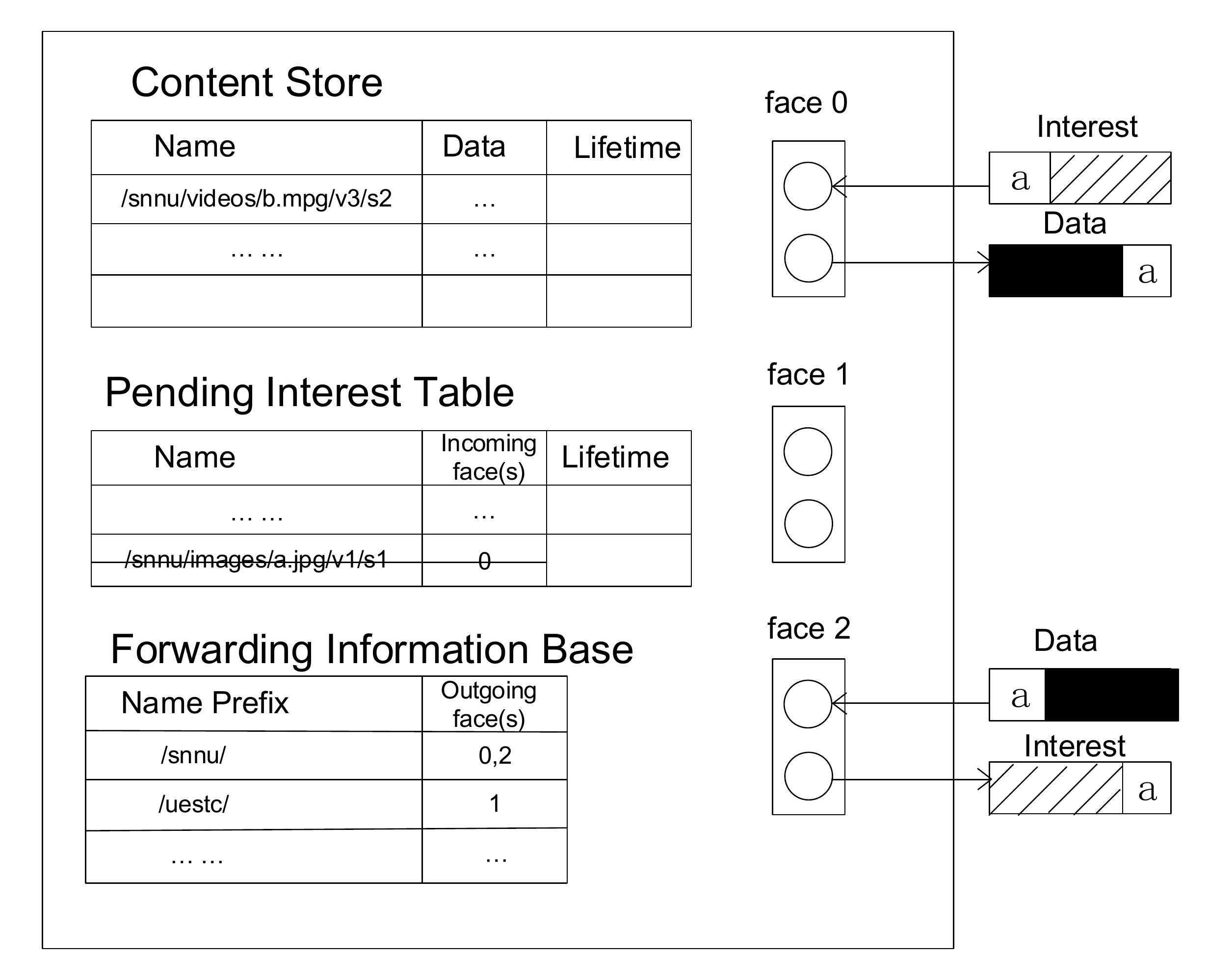}\\
  \caption{Forwarding Process of an NDN Node}\label{fig:example}
\end{figure}

Let us explain how NDN works with a concrete example shown in Fig. \ref{fig:example}. Suppose a consumer wants to fetch the content $a.jpg$ mentioned above. The consumer initializes an \emph{Interest} packet and forwards it to the network. When an NDN router gets the \emph{Interest} packet from its incoming face, the router firstly searches the entire of CS for matching data. If there exists the content, the router returns the \emph{Data} packet immediately through the face where the \emph{Interest} came and discard the \emph{Interest}. Once the CS miss caching the required data, the router checks whether the name already exists in PIT. If so, adding the face into this entry's incoming face(s) list and discarding this \emph{Interest}. Otherwise, the router looks up in FIB for the longest-prefix matching to find out where to forward the \emph{Interest}. Meanwhile, the router adds a new record in PIT including name and incoming face information. After receiving the feedback \emph{Data} packet with the exact-match name from another router, the router erases the corresponding record in PIT and forwards the \emph{Data} packet to the consumer through incoming face(s) information. Finally, the router stores a copy of the \emph{Data} packet for future reuse, thus saving the bandwidth and reducing the time consumption in data retrieval.

\section{Content-Centric Security Challenges in NDN}

NDN concentrates on information objects or contents retrieval from a network instead of the location of these contents \cite{ndn}. Thus, securing the contents themselves in NDN is more important than protecting the infrastructure of the network. In NDN, content security is a challenge since there exist a number of security concerns of NDN content including naming related attacks, caching related attacks, routing related attacks and other attacks \cite{ICNsurvey}.

\textbf{Naming related attacks.} Watchlist and sniffing attacks are two types of naming related attacks. Different from TCP/IP protocol, in which an address represents a host in the network and is not semantically related to the content, NDN names the content for facilitating data dissemination and routes the content based on the content names. This approach causes security risks as the content names are visible and semantically correlated to the content itself. Adversaries are able to monitor Internet usage with the visible content requests, which puts the NDN publisher's privacy a worse situation. Moreover, denial of service may occur as an adversary can prevent NDN users from requesting for the marked content, which leads to unanswered requests.

\textbf{Caching related attacks.} Content caching at intermediate nodes which is also referred to as in-network storage, is fundamentally necessary to support P2P data delivery in NDN at low communication cost. There are a number of benefits using caching in NDN, such as dissociating contents from their producers, avoiding single point of failure, reducing the network load and data dissemination latency, etc. NDN is built on consumer driven caching with target to delivering the closest copy available to a subscriber. Thus, NDN is vulnerable to caching related attacks including time analysis, caching pollution and bogus announcements.
Time analysis can violate users' privacy by recognizing if a user requested a content or not. Caching pollution attack states that users will be served with illegitimate content by a polluted cache, once the cache has been poisoned by an attacker. In a bogus announcement attack, an adversary sends a mass of announcement updates for content at a frequency exceeding the local content request routing convergence time to damage the caching and routing systems. As a result, the overloaded bogus contents make the subscribers fail to retrieve complete or correct content.

\textbf{Routing related attacks.}
In NDN, the content delivery relies much on asynchronous publication and subscription. Thus, some attacks such as jamming attack and time attack may come into being to violate this state consistency. There exist other routing and forwarding related attacks such as flooding attack to exhaust the resources such as memory and computational power which are employed to maintain and exchange content states. Routing related attacks can lead to denial of service, resource exhaustion, path infiltration and privacy leakage.

\textbf{Other attacks.} There are some other attacks that do not fall into the aforementioned categories, which might cause insufficient or incorrect data distribution in NDN. In packet mistreatment scenario, an adversary accesses NDN nodes to modify packets during transmission, reply the consumer plenty of times or pretend to be the publisher to create content. Unauthorized access attacks, which enable an adversary to access some content sent to legitimate users that he/she has no right to access, becomes easier as an adversary can use any available copy of a content distributed in different network locations. These attacks can lead to unauthorized access to content, congestion and denial of service etc.


\section{Protecting Content with Digital Signatures}

As the cryptographic analogue of handwritten signatures, digital signatures are a useful cryptographic primitive for authenticating an entity and validating the integrity of a message. As a fundamental security guarantee of NDN, a digital signature scheme is employed to sign the name of each NDN packet together with the packet content, which provides content integrity and origin authentication by binding the name and the data of the packet. This approach separates trust in content and in entities, which makes the globally addressable content publicly authenticated. A consumer who is the requester of some content can validate the signature and a router can also choose to check the content integrity and provenance with the signature. Moreover, named and signed content forms a solid foundation for building a number of applications such as a web browser protocol handler, a media streaming.

Finding appropriate digital signature schemes to secure NDN packet is challenging. Firstly, to sign and validate every packet requires an ominous need for high efficiency in signature generation, verification, transmission and storage. Secondly, a publisher's signatures on packets may leak his information such as his public key since this information is mandatory in signature verification. Thirdly, when a quantum computer comes into being, it can break all digital signature schemes used today. What kind of digital signature algorithms can be used in NDN in the age of quantum computers? Finally, signature validation of NDN content only reveals that the signature was generated with a particular key and thus, key management \cite{Keymanage09,Keymanage07} becomes a fundamental issue in NDN content security.

\subsection{Digital signatures}

A digital signature scheme \cite{signature} usually involves a signer and a set of verifiers. The signer begins with a key-generation algorithm to generate a key pair including a public key $pk$ and a private key $sk$.
With the key pair, a digital signature scheme allows the signer to sign a message $m$ following the \emph{Sign} algorithm to generate a digital signature on a message $m$ such that a verifier who knows $pk$ can use the corresponding \emph{Verify} algorithm to validate that $m$ is originated from the signer and keeps virgin.

\subsubsection{Cost-effective digital signatures}

Choosing cost-effective fine-grained signature schemes to sign Internet packets is a major challenge for the publisher in NDN network. NDN currently offers two choices, RSA \cite{signature} and ECDSA \cite{ECDSA}, as signature algorithms.

Let $p$ and $q$ be two odd primes and $N$ is their product.
A signer's public key is $e$ and private key is $d$ where the product of $d$ and $e$ is congruent to 1 modulo the value of Euler's phi function.
To generate a signature of message $m$, the signer firstly computes the hash value of $m$, and then the signature of $m$, that is the hash value to the power of $d$. A verifier checks if the hash value of $m$ equals to the signature value of $m$ to verify a signature.
 When one uses a 1024-bit modulus, the size of a RSA signature is 1024-bit, which might be too long to be employed in NDN content verification.

$g$ denotes a generator of a group $G$ of prime order $q$, and $h$ is an arbitrary element of $G$. The discrete logarithm of $h$ with respect to $g$ is defined as the smallest non-negative integer $x$ such that $h$ equals to the value of $g$ to the power of $x$. Discrete logarithm assumption says that it is hard to find the integer $x$ given $g$ and $h$. A number of cost-effective digital signature schemes have been proposed based on discrete logarithm assumption, and the Digital Signature Algorithm (DSA), the core of Digital Signature Standard (DSS), is among the representative signature algorithms. When one uses a 1024-bit modulus, a DSA signature is 320 bits long. ECDSA, the variants of DSA based on elliptic curve, is 320-bit long as well.

Another cost-effective digital signature is the BLS short signature from the Weil pairing \cite{signature} due to Boneh, Lynn and Shacham. BLS signature is approximately $170$ bits long while offers a security level similar to that of 320-bit DSA signature. The unforgeability of BLS signature depends on the CDH assumption on some elliptic curves over a finite field.

\subsubsection{Privacy-preserving digital signatures}
%

Digital signatures can bind content to its publisher, however, ordinary cost-effective signatures are publicly verifiable, which may leak sensitive identity information of the publisher. The privacy disclosure of publisher's identity is dangerous in NDN especially when considering censorship and monitoring since an adversary is able to identify the content from certain publishers easily from the public key explicitly stated at very NDN packet. We introduce two types of privacy-preserving digital signatures namely group signature and ring signature, to protect signature privacy, where the identity of the publisher can be hidden among a group of users.

Different from traditional digital signature schemes where only one signer is involved, group signatures \cite{Chaum91} can conceal the identity of the signer with a group of members such that the verifier can be convinced that the signer is in the group but the real identify of the signer is unknown. In a group signature, a group manager is involved to build the group, add and revoke group members.
The property of signer-ambiguity makes group signature an important building block in a variety of applications where identity privacy is a concern. Group signatures can be applied to NDN to protect publisher's identity privacy, but the presence of a trusted group manager limits group signatures to settings with collaborating publishers.


Collaboration between signers is not always possible, ring signature\cite{AOS02}, a simplified group signature, can provide ad-hoc group with no trusted group manager.
Ring signatures can provide unconditional 1-out-of-$n$ anonymity. Besides, ring signatures satisfy spontaneity and unlinkability.
The property of spontaneity of ring signature refers to the unawareness of other members being involved in the ring. Unlinkability states that two signatures generated by the same signer cannot be linked in any possible way. In ring signatures, singers can choose other users at his will to form a ring, which is more practical to implement for a NDN publisher due to no need to manage a group. Two challenges are posed if ring signatures are applied to protect NDN publisher's identity privacy: how to trace a malicious publisher for poisonous content and censorship might be still possible if using a small ring. The potential solutions are employing accountable ring signature and a bigger ring.

\subsubsection{Network coding signatures}

NDN takes advantage of the traditional ``store and forward" paradigm in the network. To increase the handling ability of the network and improve robustness against random faults in NDN network, we can integrate network coding \cite{networkcoding} with NDN, since intermediate nodes in NDN have a certain storage and computing ability.
 Homomorphic signature \cite{homomor} can be applied to NDN network with network coding.
When an interest packet comes to a NDN producer, the producer divides the corresponding data into an ordered sequence of $n$-dimensional vectors.
  Before transmitted to network, the producer generates $m$ augmented vectors by concatenating an $m$-bit long vector with every original vector $v_i$, which contains '1' only at the $i$th point and '0' at the other positions.
%
 Thus, the linear combination coefficients are included in the right-most $m$ positions. Then the augmented vectors are sent to the NDN consumer together with the producer's signatures. Routers in NDN can linearly combine the data packet produced by the same producer as well as their signatures due to the homomorphic property of the signature.
Finally, the consumer is able to retrieve the original data from sufficiently random linear combinations in a full rank matrix.

\subsubsection{Post-quantum digital signatures}
 Quantum computers can solve some computationally intractable mathematical problems efficiently, like factoring problem and discrete logarithm problem. That is, the classical digital signature algorithms used today are not secure any longer in quantum computer era and thus cannot be used in NDN network. Fortunately, there still exists some mathematical problems that cannot be efficiently solved by a quantum computer, such as the closest vector problem, the decoding problem and finding the solution of multi-variate quadratic systems. Based on these intractable problems, four types of post-quantum signature schemes, namely multivariate signatures, code-based signatures, hash-based signatures, and lattice-based signatures have received extensive attention.

\subsection{Signature implementation}

In this section, we show the implementations of the selected digital signatures, including RSA \cite{signature}, DSA \cite{signature}, BLS \cite{signature}, ECDSA \cite{ECDSA}, a group signature \cite{Chaum91} and a ring signature \cite{AOS02}.

\textbf{Environment}. All the algorithms are conducted on a desktop with 64-bit Win 10 operating system and 16.0 GB RAM. The processor is Intel(R) Core(TM) i7-7700 CPU @3.6GHz. We implement the programs in C++ projects with the help of the interfaces in Miracl library \footnote{https://certivox.org/display/EXT/MIRACL} to realize the operations in big integers and elliptic curve groups and compile the projects in VS 2010.

\textbf{Parameters}. The parameters in the implemented algorithms are fixed with the reference of ANSI X9 and Federal Information Processing Standards(FIPS 186-4, 2013)\footnote{FIPS PUB 186-4: Federal information processing standards publication. Digital Signature Standard (DSS). NIST, Gaithersburg, MD, 20899-8900, 2013. http://dx.doi.org/10.6028/NIST.FIPS.186-4. Access on July 2013.}, a golden standard of digital signatures by National Institute of Standards and Technology (NIST). NIST and ANSI X9 have set the minimum key length requirement, in which RSA and DSA are 1024 bits, and ECC is 160 bits, with the equivalent security of the corresponding symmetric block cipher a key length of 80 bits. The concrete settings are specified as follows. For RSA, we set the module $N$ 1024 bits, $p$ and $q$ 512 bits, which satisfies $|p-q|>2^{412}$. For DSA, $p$ and $q$ are selected to be 1024 bits and 160 bits respectively. For ECDSA, we employ a pseudo-random curve over prime field $GF(p)$ where $a$ is -3 and the cofactor is 1, which is one of the recommended elliptic curves by NIST for Federal Government use\footnote{https://csrc.nist.gov/projects/digital-signatures}.
For BLS, we select Miyaji-Nakabayashi-Takano (MNT) nonsupersingular curves \cite{MNT01}, in which the embedding degree is 6 and the cofactor is 2. Both group signature and ring signature are DL-based, which can share the parameters in DSA. The sizes of the group and ring are fixed as 5.

\begin{figure}[h]
  \centering
  \includegraphics[width=0.4\textwidth]{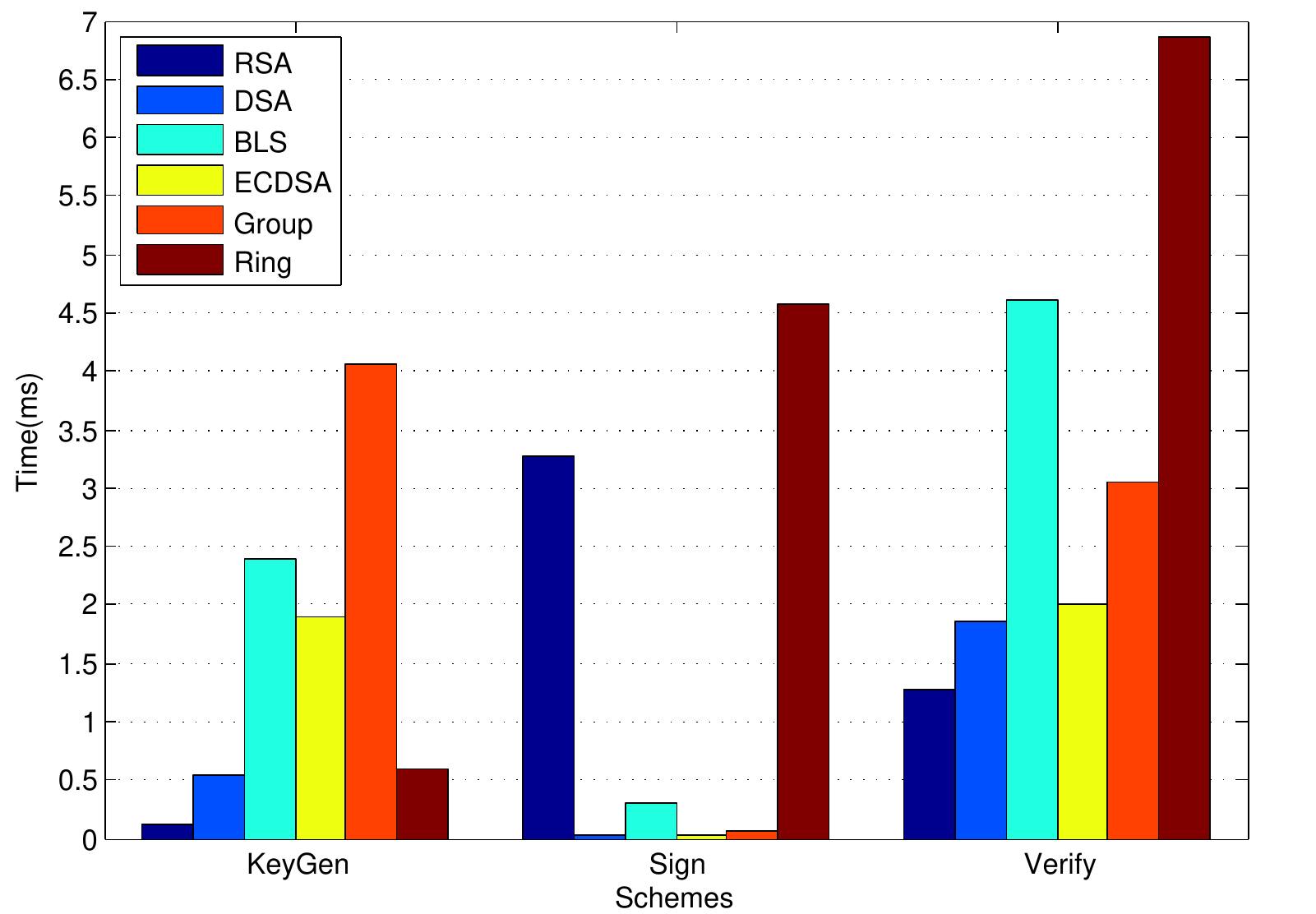}\\
  \caption{Implementation}\label{fig:implementation}
\end{figure}

\textbf{Implementation Results}. To test \emph{KeyGen}, \emph{Sign} and \emph{Verify} algorithms of each scheme, we repeat each algorithm 1,000 times and get the average running time. The result is shown in Fig \ref{fig:implementation}. In \emph{KeyGen}, the most expensive scheme is group signature with 4.059 ms time consumption, whereas ring signature, with the same number of users, only costs 0.599 ms, which is almost the same magnitude with DSA. It takes the least for RSA to generate a key pair.
As for ECC-based BLS and ECDSA, the running time is 2.400 ms and 1.890 ms respectively. In \emph{Sign} scheme, ring signature costs the most time of 4.575 ms, followed by RSA 3.274 ms. The other four schemes are all $\mu$s-level. For \emph{Verify}, the ring signature is the most time-consuming one with the performance 6.869 ms, followed by BLS 4.600 ms, group signature 3.050 ms, ECDSA 2.000 ms, DSA 1.869 ms and RSA 1.274 ms.

\subsection{Enhancing digital signatures' performance}

We introduce some techniques to enhance the performance of digital signatures from three aspects, namely speeding up the signature generation, speeding up the signature verification and reducing the bandwidth of digital signatures.

\subsubsection{Speeding up signature generation.}
Pre-computation is an important and widely used technique to make the signature generation faster. Online/offline signatures was proposed aiming to improve the signature generation by computing a signature in two phases: off-line phase (before the message to be signed is presented) and on-line phase (after the message to be signed is given).
 Shamir et al. \cite{online} presented an efficient approach by using a trapdoor hash function such as chameleon hash. With this transformation, any signature scheme is able to be transformed into a highly efficient on-line/off-line signature and the signature size does not increase too much.

\subsubsection{Speeding up signature verification.}  Verification cost of a digital signature is likely a more important factor than that of signature generation as a digital signature is generated once but may be verified many times. Batch verification and server-aided verification are two kinds of techniques to significantly reduce the signature verification time.
Batch verification \cite{batch} enables a verifier to take less time in validating multiple digital signatures simultaneously than the overall time for all the single signature verification.
Various digital signatures such as RSA, ECDSA and BLS signture have got their batch verification algorithm. Some algorithms support batch verification for the signatures from the same signer while a few algorithms allows batch verification for many different signers. Server-aided verification can speed up the signature verification by delegating a substantial part of computations to a powerful but possibly untrusted server. The computational cost of signature verification such as BLS signature can be decreased by around 70\% with server-aided verification.

\subsubsection{Reducing signature bandwidth.} The communication cost is a major issue of signature performance since signature transmission usually consumes more power than that of signature generation and verification. Signature aggregation is an important and commonly used technique to save communication bandwidth when transmitting signatures. NDN can adopt signature aggregation to achieve this type of signature compression. A publisher signs $n$ distinct packets and obtains $n$ signatures on these packets, and then aggregates all the signatures into a single signature, which can convince a consumer that the publisher signed the $n$ packets. A number of digital signatures such as RSA signature and BLS signature can be aggregated.

\subsection{Trust Management}

A valid signature on a NDN content only indicates that this signature was indeed generated with a particular key and trust management is mandatory to make this knowledge useful. Hierarchical public key infrastructure (PKI) has been built but suffers usability problems in the previous research in trust management in networks. To authenticate a public key, a digital certificate, a digitally signed statement binding the identity of an entity and his public key, is employed in PKI. However, PKI based trust management has received a lot of criticism due to its complexity of supporting certificates.

Some certificate-free trust management approaches can be potential solutions to address NDN trust management issue. We introduce four paradigms of certificate-free trust management mechanisms briefly, namely self-certified trust management, identity-based trust management, certificateless trust management, and certificate-based trust management. The trick of self-certified trust management is the public key is jointly generated by the signer and the authority such that the certificate is embedded in the public key itself. Public keys in identity-based trust management are derived directly from user identities such as email address, IP address etc while the private key is generated by a trusted third party. The elimination of certificates in identity-based trust management brings a number of benefits such as no certificate look up, no certificate management and no certificate revocation etc. The disadvantage of this approach is key escrow, that is, the trusted third party knows the private keys of all users. Certificateless and certificate-based trust management can reduce the trust levels of the trusted third party in identity-based trust management as in these systems, the full public key of a user is based on user-generated public key component and the user's identity while the full private key is created from user-generated private key component and the private key component generated by the semi-trusted third party. Recently, blockchain based PKI attracted some attention and Emercoin \footnote{https://en.wikipedia.org/wiki/Emercoin}, a blockchain-based cryptocurrency that storages different public key types has been developed.

\section{Conclusion}
In this article, we investigated content integrity and provenance issue in NDN and concluded that digital signature is a fundamental and essential tool to address this issue. We introduced diverse digital signatures to achieve content integrity and origin authentication in NDN for various scenarios. We also proposed different techniques to speed up the signature generation and verification to make signatures practical in NDN. Several types of certificate-free trust management was introduced as well to manage the keys of signature schemes. Our future research is to deploy these signatures and techniques in the real-world NDN environment.

{\textbf{Acknowledgments.}}
This work was supported by the National Key R\&D Program of China (No.2017YFB0802000), the NSFC Research Fund for International Young Scientists (61750110528), the NSFC grant (61772326, 61572303), National Cryptography Development Fund (MMJJ20170216), and the Fundamental Research Funds for the Central Universities (GK201702004).


\textsf{Yong Yu} is currently a Professor of Shaanxi Normal University, China. His research interest is blockchain. He is an Associate Editor of Soft Computing.

\textsf{Yannan Li} is a Ph.D. candidate of University of Wollongong, Australia. Her research interest is blockchain.

\textsf{Xiaojiang Du} is currently a professor of Temple University. He has published over 200 papers and has been awarded more than \$5M research grants. His research interests are security, systems, and computer networks. He is a Senior Member of the IEEE.

\textsf{Ruonan Chen} is currently a master candidate of Shaanxi Normal University. Her research interest is digital signature.

\textsf{Bo Yang} is a Professor of Shaanxi Normal University. His research interest is cryptography.
%
%
%
%
%
%

\vfill\eject

\end{document}